\documentclass[aps,prl,twocolumn,groupedaddress]{revtex4}
\usepackage{graphicx}

\begin{document}

\newcommand{\Ln}{\ensuremath{L_{n}}}
\newcommand{\Ls}{\ensuremath{L_{s}}}
\newcommand{\EF}{\ensuremath{E_{F}}}
\newcommand{\ETh}{\ensuremath{E_{Th}}}
\newcommand{\Eg}{\ensuremath{E_{g}}}
\newcommand{\EG}{\ensuremath{E_{gap}}}
\newcommand{\Ei}{\ensuremath{E_{1}}}
\newcommand{\Eii}{\ensuremath{E_{2}}}
\newcommand{\Dn}{\ensuremath{D_{n}}}
\newcommand{\Ds}{\ensuremath{D_{s}}}
\newcommand{\deln}{\ensuremath{\xi_{n}}}
\newcommand{\dels}{\ensuremath{\xi_{s}}}
\newcommand{\xin}{\ensuremath{\xi_{n}}}

\title{Anomalous density of states in a metallic film in proximity with a superconductor}

\author{A. K. Gupta, L. Cr\'{e}tinon, N. Moussy, B. Pannetier and H. Courtois}
\affiliation{Centre de Recherches sur les Tr\`es Basses Temp\'eratures - C.N.R.S. associ\'e \`a l'Universit\'e Joseph Fourier, 25 Avenue des Martyrs, 38042 Grenoble, France.}

\date{\today}

\begin{abstract}
We investigated the local electronic density of states in superconductor-normal metal (Nb-Au) bilayers using a very low temperature (60 mK) STM. High resolution tunneling spectra measured on the normal metal (Au) surface show a clear proximity effect with an energy gap of reduced amplitude compared to the bulk superconductor (Nb) gap.  Within this mini-gap, the density of states does not reach zero and shows clear sub-gap features. We show that the experimental spectra cannot be described with the well-established Usadel equations from the quasi-classical theory.
\end{abstract}


\maketitle

At the contact with a superconductor (S), the Andreev reflection of the electrons from a Normal metal (N) locally modifies the N metal electronic properties, including the local density of states (LDOS) \cite{Revue}.  The precise LDOS spectra depend on the N-S structure geometry, in particular the N metal length, and on the electron transport regime.  In a diffusive N-S junction with a N metal shorter than the phase coherence length, one expects a fully opened mini-gap, which remains smaller than the superconductor's energy gap $\Delta$ \cite{Belzig-PRB}.  In a larger N metal, the LDOS shows, within a pseudo-gap, a linear evolution with the energy, reaching zero precisely at the Fermi level.  In the ballistic regime, the same distinction holds between a chaotic (mini-gap) and an integrable (pseudo-gap) cavity \cite{Beenakker-EPL}.  In general terms, a mini-gap shows up if every electronic state at the Fermi level can couple to the S interface while maintaining quantum phase coherence.  The order of magnitude of the energy gap $E_{g}$ is $\hbar/\tau_{AR}$, where $\tau_{AR}$ is the characteristic diffusion time before an electron experiences an Andreev reflection.  Here, we shall consider the case of diffusive N and S metals brought in contact through a highly transparent interface.  Then $\tau_{AR} \simeq L_{n}^2/D_{n}$ and the predicted gap $E_{g}$ is about the Thouless energy ${\hbar}{\Dn}/{\Ln}^{2}$, where {\Ln} and $D_{n}$ are the length and the diffusion constant of the normal metal.

Recently, the LDOS of lateral S-N (Nb-Au) structures was probed with solid tunnel junctions \cite{Gueron-PRL} and (very) low temperature STM \cite{Moussy-EPL,Vinet-PRB}.  These studies focused on the pseudo-gap regime in long N metals.  Also, a mini-gap was observed in a very thin (20 nm) Au layer on top of a Nb dot \cite{Moussy-EPL}.  A good agreement with the quasi-classical theory based on the Usadel equations \cite{Belzig-PRB} and with the Bogoliubov-de Gennes equations \cite{Halterman-PRB} was obtained.  In the mini-gap regime, a NbSe$_{2}$ crystal covered with a varying thickness of Au was studied by STM at 2.5 K \cite{Dynes-PRL} but the temperature did not allow the observation of a fully open mini-gap.  Therefore, the mini-gap regime remains to be investigated, in particular its evolution with the normal metal size and the possible presence of states within the gap.  A clear distinction between a fully open gap and a pseudo-gap requires a high energy resolution $k_{B}T \ll \Delta$, which can be achieved only at very low temperatures ($T \ll 1K$) and with a large gap $\Delta$.

In this paper, we present measurements of the local density of states at the N metal surface of S-N bilayers with a varying N metal length. We used a very low temperature STM as the local tunneling probe, which permitted us to avoid the averaging over the surface inhomogeneities.  The achieved high energy resolution enabled us to unveil a new phenomenon, namely a non-zero density of states appearing in the vicinity of the Fermi level together with clear sub-gap structures, and an anomalous mini-gap amplitude dependence with the N metal length.  We demonstrate that these new features cannot be explained within the well-established quasi-classical theory.

\begin{figure}
\includegraphics[scale=0.94]{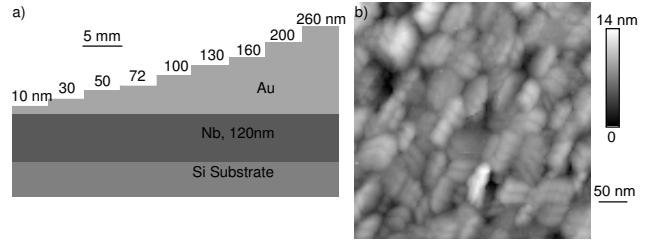}
\vspace{-0.1 cm}
\caption{\label{fig:scheme} a) Schematic cross section of the full Nb-Au bilayers sample.  b) STM image (410 $\times$ 410 nm$^2$) at 100 mK of the sample with a Au thickness of 72 nm.  The rms roughness for this image is 3.4 nm.}
\end{figure}

We fabricated simultaneously a series of Nb-Au bilayers, with a fixed Nb thickness and a varying Au thickness, on a single (6$\times$40 mm$^{2}$) Si substrate.  The Nb layer thickness was chosen to be significantly larger than the superconducting coherence length in order to avoid any effects due to its finite value.  For varying the Au thickness, another Si wafer was used as a mask and moved in situ above the substrate.  After depositing the 120 nm Nb film, the Au film (from 10 to 260 nm) was deposited within 15 minutes at a pressure below 10$^{-8}$ mbar.  These conditions minimize the interface contamination and should preserve the best Nb-Au interface transparency.  The full Si wafer with the bilayer films thus obtained (see Fig.  \ref{fig:scheme}a) was cleaved in air to separate the different samples.  Individual Au (260 nm) and Nb (120 nm) layers were characterized by transport measurements which are summarized in Table \ref{tab:trans-para}.

\begin{figure}[t]
\includegraphics{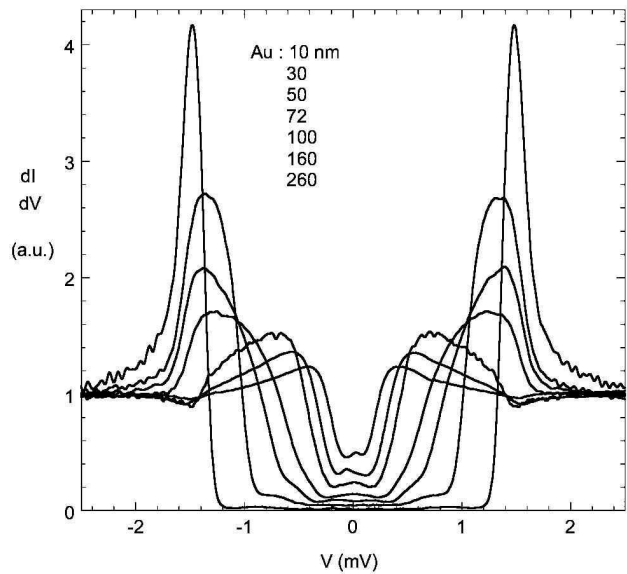}
\vspace{-0.2 cm}
\caption{\label{fig:spectra}Tunneling density of states measured at 60 mK at the Au surface of Nb-Au bilayer samples with a varying Au thickness \Ln.  Data from the 130 and 200 nm samples are not shown for ease of reading.}
\end{figure}

Each bilayer sample was cooled separately in a STM working at 60 mK in an upside-down dilution fridge \cite{Moussy-RSI}.  A fresh-cut Pt-Ir wire was used as the STM tip.  A typical STM image (of the 72 nm sample) acquired at very low temperature (100 mK) is shown in Fig. \ref{fig:scheme}b.  We observe a polycrystalline structure with a grain size of 30 - 50 nm.  The same grain size was observed in every sample, except in the thinnest one (10 nm). Moreover, this value is consistent with the measured elastic mean free path $l_{e,n}$ in the 260 nm sample.  Our Au films are thus in the diffusive regime, except for the smallest thicknesses where ballistic effects may occur.  On every sample, we acquired series of $I(V)$ tunnel characteristics at a tunnel resistance of 5 to 12 $M\Omega$ and at numerous places in order to check reproducibility.  The spectra did not depend on the tunnel resistance value and were very reproducible over a sample surface with minor details (in particular small asymmetries at zero bias) slightly evolving, and also punctual defects on a few locations. The $I(V)$ data were numerically differentiated to obtain the differential conductance $dI/dV(V)$, which gives the LDOS at the energy eV with an accuracy limited by the thermal smearing.  A selection of the tunneling spectra is shown in Fig. \ref{fig:spectra}. The spectra are flat at large bias voltages ($\left|V\right| \gg$ 2.5 meV) and have been normalized as 1 there.

\begin{table}[t]
\caption{\label{tab:trans-para}Transport properties of the S=Nb (120 nm) and N=Au (260 nm) films at 10 K. The RRR is the ratio of the room-temperature resistivity with the 10 K residual resistivity.}
\begin{ruledtabular}
\begin{tabular}{cccccc}
 &$\rho$ ($\mu \Omega.cm$)&RRR&$l_e$ (nm)&D ($cm^2$/s)&$\xi$ (nm)\\
 \hline
Nb&16.2&3.7&5.4&24.7&23.2 \\
Au&2.34&4.7&36.1&169&60.8 \\
\end{tabular}
\end{ruledtabular}
\end{table}

For the bilayer with the smallest Au thickness (10 nm), the spectrum qualitatively resembles a BCS spectrum with a gap amplitude very close to the expected bulk Nb gap : $\Delta \sim 1.5$ meV. For larger thicknesses, the spectra show a mini-gap that reduces in width with increasing Au thickness.  There seems to be a crossover from one type of spectra to another at a Au thickness between 72 and 100 nm. For small thicknesses, there is a sharp rise to a peak in the LDOS at $\Delta$, and a relatively slow decrease as we go towards zero bias. For large thicknesses, there is a small dip at $\Delta$, the LDOS then rises slowly to a peak and falls off rapidly inside the mini-gap. This dip at $\Delta$ actually implies that the superconducting gap in the bulk of Nb is not significantly influenced by a large thickness of Au.

Inside the mini-gap, the density of states is strikingly different from zero and features a few sub-gap structures.  The observed sub-gap features do not scale with the Nb energy gap $\Delta$, as their energy position evolves with the normal metal length {\Ln}. At the Fermi level (V = 0), the mean LDOS monotonously increases with the Au thickness.  This effect is therefore not a ballistic one which would be restricted to the smaller thicknesses.

We checked the energy resolution of our STM by measuring the LDOS at the surface of plain Nb and Al films and fitting the data with a BCS density of states without any depairing parameter but with a thermal smearing.  The fit was very good and gave an effective temperature of 210 mK which corresponds to an energy resolution of 36 $\mu$eV \cite{Moussy-RSI}.  This proven high resolution rules out the thermal smearing as a cause of the large (up to 50\%) zero bias conductance observed here.

We compared our data to the predictions of the quasi-classical theory in the diffusive regime \cite{Golubov-JLTP,Belzig-PRB}, by solving the Usadel equations with the help of a numerical code from W. Belzig et al.  \cite{Belzig-PRB}.  The input parameters for this calculation are the lengths $L_{n,s}$ of N and S metals in units of the related coherence lengths $\xi_{n,s}=\sqrt{\hbar D_{n,s}/2\Delta}$, the specific interface resistance $r_{B}$, the inelastic $\gamma_{in}$ and spin-flip $\gamma_{sf}$ scattering rates in the N and S metals, the electronic properties mismatch parameter $\gamma = {\rho}_{s}\dels/ {\rho}_{n}\deln$ $(\propto \sqrt{l_{e,n}/l_{e,s}})$.  Here $\rho_{n,s}$ is the normal-state resistivity in N or S. We considered the interface transparency as perfect and we took into account a thermal smearing with an effective temperature of 210 mK.

\begin{figure} [t]
\includegraphics{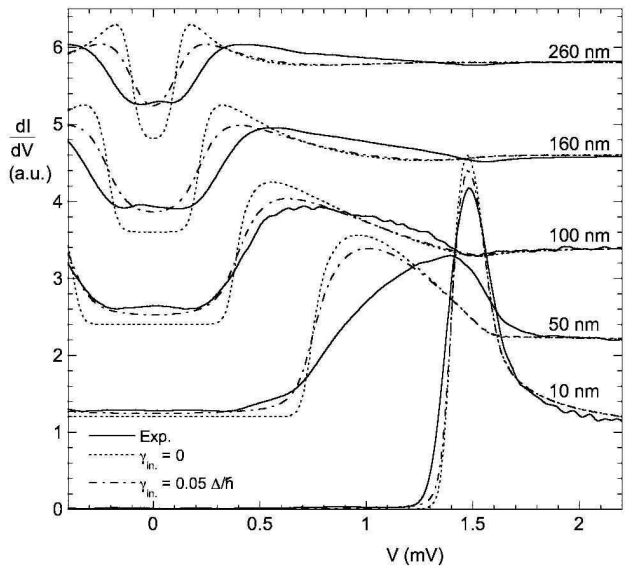}
\vspace{-0.2 cm}
\caption{\label{fig:detail} Comparison of the experimental data for {\Ln}= 10 ; 50 ; 100 ; 160 and 260 nm (full line) with the calculated spectra.  The calculation parameters are $L_{s}/\xi_{s}$ = 5.17, $\Delta$ = 1.57 meV, $T$ = 210 mK, $\gamma$ = 0.6, $r_{B}$ = 0, and $\xi_{n}$ = 60.8 nm.  The dotted line curve stands for $\gamma_{in}$ = 0 and the dashed line curve for $\gamma_{in}$ = 0.05 $\Delta/\hbar$. The curves have been evenly shifted for clarity.}
\end{figure}

We adjusted the calculation parameters to fit the tunneling spectra.  Fig.  \ref{fig:detail} displays a comparison of experimental spectra with two sets of calculated curves.  We first assumed that the inelastic and spin-flip scattering lengths are much larger than the Au layer thickness, so that we can neglect these processes (Fig.  \ref{fig:detail} dotted lines).  In order to recover the observed LDOS peak amplitude and position, we had to assume a mismatch parameter $\gamma$ value of 0.6 instead of 2.6 as estimated from the transport measurements, keeping the other parameters matching precisely the measured values.  This discrepancy may be related to the Nb-Au interface roughness which is expected to affect the Andreev reflection rate, and thus the induced superconductivity in the Au layer \cite{Pilgram-PRB}.  The overall spectral shapes are qualitatively reproduced, except for the peak of the 50 nm spectra which is noticeably different from the experimental data.  At large {\Ln}, the predicted mini-gap is clearly smaller than the one observed experimentally.  Moreover, we always obtain a fully opened gap with a zero LDOS at the Fermi level.  This is in clear disagreement with our experimental results, where we always get a non-zero conductance at zero bias.  The dependence of the mean free path with the Au layer thickness {\Ln} \cite{Fuchs-PCPS} was taken into account through the related variation of the characteristic length $\xi_{n}$ and of the mismatch parameter $\gamma$.  This modified only slightly the calculated spectra in the regime \Ln $<$ 100 nm and did not improve the fit.

\begin{figure}[t]
\includegraphics{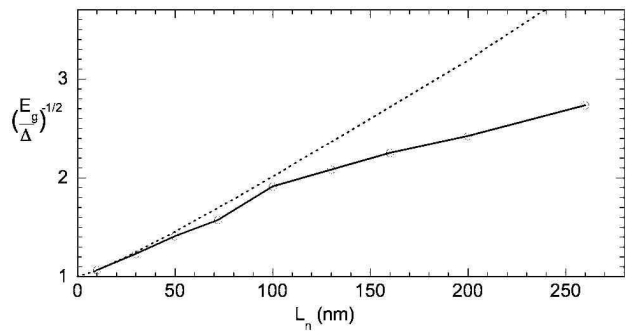}
\vspace{-0.2 cm}
\caption{\label{fig:gapvar} Variation of the quantity $(E_{g}/\Delta)^{-1/2}$ with the N metal length \Ln ; the mini-gap $E_{g}$ being defined by the inflection point in $dI/dV(V)$.  The dotted line shows the mini-gap calculated from Fig.  \ref{fig:detail} parameters ($\gamma_{in}=0$).}
\end{figure}

With the hope to obtain a better fit, we included a non-zero inelastic scattering rate in the calculation parameters (Fig.  \ref{fig:detail} dashed lines).  We achieved a good agreement of the calculated LDOS with the measured one, but only at the precise Fermi level, by choosing $\gamma_{in} = 0.05 \Delta/\hbar$.  This value corresponds to an inelastic mean free path of about 370 nm, which is significantly smaller than the expected value of about 2 $\mu m$.  At large thicknesses, the calculated mini-gap remains smaller than in the experiment.  Fitting the mini-gap width requires a significant modification of the calculation parameters such that the calculated LDOS close to the Fermi level also diminishes.  This makes the disagreement between the experimental data and calculated LDOS even stronger for a fixed value of $\gamma_{in}$.  We also tried to include a spin-flip scattering \cite{Gueron-PRL,Vinet-PRB} or a small interface resistance in the calculation, but it did not improve the fit.

From the experimental tunneling spectra, we extracted the mini-gap $E_{g}$ as the energy at the inflection point, i.e., with a maximum slope $d^{2}I/d^{2}V$.  The quantity $(E_{g}/\Delta)^{-1/2}$ is plotted in Fig.  \ref{fig:gapvar} as a function of the N metal length $L_{n}$, so that a Thouless energy scaling should appear as a linear behavior.  The calculated mini-gap is close to this scaling : ${\Eg} = 1.3{\times} 10^{4}/(76+{\Ln})^{2}$ meV, $\Ln$ being in nm units.  As naively expected, the numerator value is of the order of $\hbar D_{n}= 1.09{\times}10^{4}$meV$.$nm$^{2}$, based on Table 1 data.  The extra length appearing in the numerator expresses the fact that the Andreev reflection happens in the superconductor over this length scale.  It also makes $E_{g}$ extrapolate to $\Delta$ as {\Ln} goes to zero.  The experimental data follows this behavior in the small thickness regime \Ln $<$ 100 nm.  At large {\Ln}, the discrepancy between theory and experiment is striking.

\begin{figure} [t]
\includegraphics{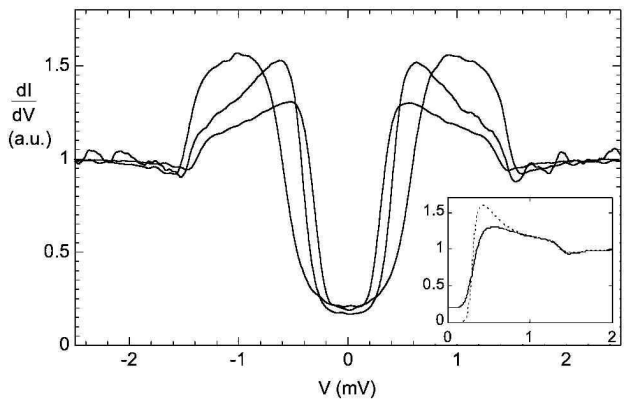}
\vspace{-0.2 cm}
\caption{\label{fig:ArEtch} Spectra acquired at the surface of Nb-Au bilayers with fixed Nb and Au lengths \Ls = 60 nm, \Ln = 20 nm and an Ar etched Nb-Au interface.  From large to small mini-gap, the etch times are 30 s, 1 min.  30 s and 3 min.  Inset : fit of the 3 min.  etched sample data with the Usadel equations solution with $L_{s}/\xi_{s}$ = 2.5, $L_{n}/\xi_{n}$ = 0.7, $\Delta$ = 1.47 meV, $T$ = 210 mK, $r_{B}$ = 6 $m\Omega$.$cm^{2}$, $\gamma$ = 1.1, $\gamma_{in}$ = 0.}
\end{figure}

Up to now, we described samples with an exceptional interface transparency.  We have also fabricated similar Nb-Au samples but with a degraded interface, thanks to an Ar etch of the Nb surface performed just before a 20 nm Au deposition.  Different Ar etch times were used, keeping Ar pressure (3$.  10^{-4}$mbar) and beam voltage (100 eV) fixed.  Compared to Fig.  \ref{fig:spectra} data at similar thickness, Fig.  \ref{fig:ArEtch} tunneling spectra show a drastic change in the LDOS. A mini-gap is observed, with a decreasing amplitude as the etching time is increased.  We interpret this behavior as due to a decrease of the interface transparency induced by the surface milling.  This enhances the contribution of specular reflection of electrons as compared to the Andreev reflection, which makes the time $\tau_{AR}$ increase and the mini-gap $E_{g}$ decrease.  Interestingly, these spectra also show a non-zero density of states at the Fermi level, which seems insensitive to the variation of the interface transparency.  Although we were able to qualitatively fit the mini-gap width and the peak profile with the Usadel solutions by taking into account a non-zero interface resistance (see Fig.  \ref{fig:ArEtch} inset), it was again impossible to fit the whole spectra, especially the non-zero LDOS near the Fermi level.  A moderate interface transparency thus causes an anomalous LDOS, in a similar way as a large N metal length.

The observed LDOS, especially its amplitude at the Fermi level and the mini-gap width evolution, cannot be described by the quasi-classical theory, even after assuming an extremely strong phase-breaking scattering.  In the diffusive regime of relevance here, neither the angular dependence of the Andreev reflection at an imperfectly transparent N-S interface \cite{Mortensen-PRB} nor the presence of surface electron states \cite{Everson-JVST} should play a role.  In terms of electron trajectories, a finite LDOS at the Fermi level means that a significant fraction of the electron states present at the Au surface do not Andreev-reflect at the Nb-Au interface on the time scale of the phase coherence time.

In our opinion, this new effect was not observed before because previous studies either did not feature the relevant thickness range and a confined geometry where the mini-gap enables a clear observation of additional low-energy single particle states \cite{Moussy-EPL,Vinet-PRB}, or were carried out at a higher temperature with insufficient energy resolution in the measured LDOS \cite{Dynes-PRL}.  Still, our observations are somewhat reminiscent of some unexplained features reported on the LDOS of presumably quite disordered 20 nm N metal ridges lying over a Nb surface \cite{Vinet-PRB}.  There also, a clear filling of the LDOS at the Fermi level was observed in the absence of any mini-gap width evolution.  Let us point out that both this and our systems are far from the strongly disordered regime where a non-zero LDOS for all energies is predicted \cite{Feigelman-PRL}.  Rather, our films have the typical polycrystalline structure of a weakly disordered thin film.  The elastic diffusion in the N metal is not homogenous as in an amorphous metal but controlled by the interfaces between ballistic grains.  The possible confinement of electron states within a grain can have serious effects on the LDOS in the proximity superconductivity regime of interest here.  A significant fraction of electron states may be sufficiently decoupled from the N-S interface to avoid any Andreev reflection.  These states would thus remain insensitive to the superconductivity in the S layer and contribute fully to the LDOS.

In conclusion, we investigated with a high energy resolution the LDOS at the N metal surface of diffusive S-N (Nb-Au) bilayers and compared to the theoretical predictions.  We uncovered an anomalous mini-gap width evolution and a non-zero LDOS near the Fermi level that increases as the N metal thickness is increased.  The latter behavior is also observed at small thickness when the interface transparency is reduced on purpose.  This new effect cannot be described within the quasi-classical theory.  We suggest that the granular structure of the Au layer enables the decoupling of some electron states from the Andreev reflection at the interface, which results in a non-zero LDOS inside the mini-gap.

We thank W. Belzig for sharing his numerical code, M. Feigelman for discussions and the ACI "Nanostructures" for financial support.

\end{document}